\input harvmac
\Title{\vbox{\baselineskip12pt\hbox{{\rm CERN-TH}/2001-104
}}}
{\vbox{\centerline
{Duality and Confinement in Massive} \vskip2pt\centerline{
Antisymmetric Tensor Gauge Theories }}}
\centerline{ M. C. Diamantini\footnote{*}
{Supported by a Swiss National Science Foundation fellowship.}.}
\medskip
\centerline{CERN Theory Division, CH-1211 Geneva 23,
Switzerland}

\vskip .3in
\centerline{Abstract}

We extend the duality between massive and topologically massive antisymmetric
tensor gauge theories in arbitrary space-time dimensions 
to include topological defects. We show explicitly
 that the condensation of these
defects leads, in 4 dimensions, to  confinement  of electric strings in the two dual models.
The dual phase, in which magnetic strings are confined is absent.
The presence of the confinement phase explicitely found in the 4-dimensional
case, is generalized, using 
duality arguments, to arbitrary space-time dimensions.

\Date{$\eqalign{&{\rm CERN-TH}/2001-104\cr &{\rm April}\ 2001\cr}$}

\vfill
\eject

\newsec{Introduction}

Antisymmetric tensor gauge theories have attracted much interest
in  recent years  \ref\kal{V.I.
Ogievetsky and I.V. Polubarinov, Sov. J. Nucl. Phys. 4 (1967) 156;
M. Kalb and P. Ramond, Phys. Rev. D9 (1974) 2273; Y.
Nambu, Phys. Rep. 23 (1976) 250; D.Z. Freedman and P.K. Townsend, Nucl. Phys. B177
(1981) 282.}, \ref\tei{ R.I. Nepomechie,
Phys. Rev. D31 (1984) 1921; C. Teitelboim, Phys. Lett. B167 (1986)
63; C. Teitelboim, Phys. Lett. B167 (1986) 69.}, \ref\orl{R. Savit, 
Phys. Rev. Lett. 39 (1977) 55; P. Orland,
Nucl. Phys. B205[FS5] (1982) 107.For a review see: R. Savit,
Rev. Mod. Phys. 52 (1980) 453.}, \ref\aur{A. Aurilia, Y. Takahashi and P.K.
Townsend, Phys. Lett. B136 (1984) 38; A. Aurilia, F. Legovini and E.
Spallucci, Phys. Lett. B264 (1991) 69.} 
\ref\pin{F. Quevedo and
C. A. Trugenberger, Nucl. Phys. B501 (1997) 143.}, since they arise in constructing 
gauge theories of elementary extended  objects (strings, membranes,...
): an antisymmetric tensor of rank $(p+1)$\ couples to elementary 
$p$-branes, a natural generalization of the coupling of the
vector potential one-form in Maxwell theory 
to elementary point-particles (0-branes). 
Antisymmetric tensors also appear naturally
in effective field theories for the low-energy dynamics of strings
and in supersymmetric theories \ref\gre{M. Green, J. Schwarz and E.
Witten, ``Superstring Theory'', Cambridge University Press, Cambridge
(1987).}, where they play an important role in the realization of various
dualities among string theories \ref\fer{For a review see: J. Polchinski, Rev.
Mod. Phys. 68 (1996) 1, and references therein.}. 
General dualities between different phases of antisymmetric tensor field theories were
first established in \pin. However, topological terms were not considered there.

The study of dualities is becoming more and more important due to recent
developments in string theory, where it was shown that inequivalent vacua are 
related by dualities based on the existence of extended objects,
 the D-branes \fer.
Establishing a duality means that one has  two equivalent descriptions
of the same theory in terms
of different fields. This is normally very useful because,
typically, duality exchanges the coupling constant $e\rightarrow {1\over e}$:
strong and weak coupling are interchanged, opening up the possibility of doing
perturbative calculations in all regimes of the coupling constant.

In this paper we will concentrate on  Abelian gauge invariant 
formulations for Massive Gauge Theories (MGTs) and
Topologically Massive Gauge Theories (TMGTs) for antisymmetric tensor fields.

As an alternative to the Higgs mechanism for gauge invariant masses we will
consider a St\"uckelberg-like formulation \ref\ecg{E.C.G. St\"uckelberg, Helv.
Phys. Acta 11 (1938) 225.}, and a topological coupling, the BF term, that is a
generalization to arbitrary space-time dimensions of the Chern-Simons term.
These two models have been shown to be dual in \ref\spa{E. Harikumar
and M. Sivakumar, Phys. Rev. D57 (1998), 3794; A. Smailagic and E. Spallucci,
Phys. Rev. D61 (2000) 067701.}, in case of non-compact gauge symmetry group.

The MGTs we will consider are described by the action  \pin\ 
\ref\cri{M.C. Diamantini, Phys. Lett. B388
(1996) 273.}

\eqn\opb{\eqalign{ S = &\int {(-1)^{p+1}\over g^2} dB_{p+1} \wedge
*dB_{p+1} + {(-1)^p e^2 \over 4} \left( \tilde m B_{p+1} +{1 \over e} d A_p \right) 
\wedge * \left(\tilde m B_{p+1} +{1 \over e} d A_p \right) \cr 
&+ j \left (\tilde m B_{p+1} + {1\over e} d A_p \right)
 \wedge * J_{p+1}\ , \cr }}
where $A_p$\ is an antisymmetric tensor of rank $p$, $e$\ is a
dimensionless coupling constant, $\tilde m$ is a mass parameter and 
$J_{p+1}$ is a current of $p$-branes 

\eqn\ccp{J^{\mu_1...\mu_{p+1}} (x) = \int 
\delta^{d+1}(x - y(\sigma))\ dy^{\mu_1} \wedge...\wedge dy^{\mu_{p+1}}
\ .}
Here $y(\sigma)$\ are the coordinates of the world-volume of the
$p$-branes.
$S$ is invariant under the combined gauge transformation:

\eqn\cgs{\eqalign{ B_{p+1} \quad &\rightarrow B_{p+1} + d\Lambda_p\ ,
\cr  A_p \quad &\rightarrow A_p - e \tilde m\Lambda_p\ . \cr }}
Since the term $\left ( B_{p+1} + {1\over \tilde m e} d A_p \right)$ is itself 
gauge-invariant, the current $J_{p+1}$ does not need to be
conserved, and we can couple the theory to open $p$-branes.
Up to a gauge transformation, we can rewrite \opb\ as a
generalized version of the Proca Lagrangian for an
antisymmetric tensor field $\bar B_{p+1} = \tilde m B_{p+1} + {1 \over e}\
d A_p$:

\eqn\sla{\eqalign{ S = &\int {(-1)^{p+1}\over g^2} d \bar B_{p+1} \wedge
* d \bar B_{p+1} + (-1)^p \tilde m^2 \bar B_{p+1}\wedge * \bar B_{p+1} \cr
  &+ j \bar B_{p+1} \wedge * J_{p+1}\ .\cr }}
In this formulation the higher-rank tensor $B_{p+1}$ has ``eaten'' the
tensor $A_p$, which is a generalization of the familiar
Higgs-St\"uckelberg mechanism for vector fields.

The model we will consider for TMGTs is a modified form,
generalized to arbitrary dimensions, of the mechanism proposed in \ref\jac{S.
Deser, R. Jackiw and S. Templeton, Ann. Phys. 140 (1982) 372.}\ in the contest
of 3-dimensional QED, where the photon acquires a mass due to the presence of a
topological term, the Chern-Simons action.
In our case the topological term will be a BF term of the form $B_{p+1}\wedge
d \tilde A_p$,
with an action given by:

\eqn\tmm{\eqalign{\tilde S = &\int {(-1)^{p+1}\over g^2} dB_{p+1} \wedge
*dB_{p+1} + (-1)^p \tilde m \left( B_{p+1} 
\wedge  d \tilde A_{d-p-2} \right) \cr
 &{(-1)^p \over e^2} \left( d \tilde A_{d-p-2} \wedge^* d \tilde
  A_{d-p-2} \right) 
+ j B_{p+1} \wedge * J_{p+1} + \phi \tilde A_p \wedge *\phi_p\ , \cr }}
where $J_{p+1}$ is a current of closed $p$-branes and $\phi_p$ is a current of
closed $(p-1)$-branes. In this way it is possible to obtain a gauge invariant
massive gauge theory for the $B_{p+1}$ and the $\tilde A_{d-p-2}$ form without a Higgs field,
as it has been shown in
\ref\abl{T.J. Allen, M.J. Bowick and A. Lahiri, Mod. Phys. Lett. A6 (1991)
610; R.R. Landim and C.A.S. Almeida, hep-th/0010050}.
The action \tmm\ is gauge invariant under the two independent gauge
trasformations:

\eqn\igs{\eqalign{ B_{p+1} \quad &\rightarrow B_{p+1} + d\Lambda_p\ ,
\cr \tilde A_{d-p-2} \quad &\rightarrow \tilde A_{d-p-2} + d 
\tilde \Lambda_{d-p-3}\ . \cr }}

In \spa\ it has been shown that these two models can be obtained starting from
the master action:

\eqn\mnc{\eqalign{ S_{\rm M} = &\int {(-1)^{p-1}\over g^2}  
dB_{p+1} \wedge^*
dB_{p+1}  + H_{d-p-1} \wedge \left({1 \over e} dA_p + \tilde m B_{p+1}  \right) 
 \cr &+ {(-1)^{d-p-2}\over e^2} H_{d-p-1}\wedge^*H_{d-p-1} \ .\cr}}
Integrating over $H_{d-p-1}$ leads to \opb, while integrating over the
Lagrange multiplier $A_p$ gives \tmm.

If the antisymmetric tensors are compact variables, extended objects can also
appear as topological defects in the underlying gauge theory:
they are $({\rm d}-p-2)$-dimensional extended objects
representing the world-hypervolumes of $({\rm d}-p-3)$-branes
(instantons are ($-1$)-branes). 
The presence of topological defects can lead to modifications of the
infrared perturbative behaviour: their condensation (or lack of it) can drastically
change the phase structure of the theories \ref\pbo{For a review see: A.M.
Polyakov, ``Gauge Fields and Strings'', Harwood Academic Publishers, Chur
(1987).}.

It has been shown in \pin\  that the
condensation of generic $p$-branes interpolates between
massless and massive antisymmetric tensor field theories of
different rank.
The appearance of antisymmetric tensors of higher rank in the phase in
which the $p$-branes condense is related (in the case of rank two)
to Polyakov's confining string mechanism\ref\polb{A. M. Polyakov,
Nucl. Phys. B486 (1997) 23; M.C. Diamantini, F. Quevedo and C.A. Trugenberger,
Phys. Lett. B396 (1997) 115; M.C. Diamantini and C.A. Trugenberger, Phys. Lett.
B421 (1998) 196; M.C. Diamantini and C.A. Trugenberger, Nucl. Phys. 
B531 (1998) 151. }.

In this paper we will not address the problem of establishing if topological
defects indeed condense and for which regimes of the coupling constants, but we
will concentrate, instead, in studying the nature of the phases in which a finite
condensate of topological defects exists.
In what follows we will extend the duality established in \spa\ to include
topological defects and we will analyze the effects of the condensation of these
topological defects in the two dual theories.

\newsec{Duality with Topological Defects}

To study the duality between massive and topologically massive antisymmetric
tensor gauge theories in case of a compact gauge symmetry, we will follow the approach of
\pin, and treat the topological defects explicitly.
They will be represented by singular forms $t_p$ such that
\eqn\tdf{ *t_p = V_{d-p}\ ,\quad 
V^{\mu_1...\mu_{d-p}}_{d-p} =  \int \delta^{d}\left( x -
\tilde y(\tilde \sigma) \right) d \tilde y^{\mu_1} \wedge ...\wedge
d \tilde y^{\mu_{d-p}} \ ,}
with $\tilde y^{\mu} (\tilde \sigma^{\nu_1},...,\sigma^{\nu_{d-p}})$
an open hypervolume describing the generalization to
higher-dimensional topological defects of the Dirac string. The
boundary of this hypervolume describes the world-hypersurface of the
topological defects.

In order to make both the combined gauge symmetry \cgs\ of the MGTs and the two
independent gauge symmetries \igs\ of the TMGTs compact we need to introduce
three differents types of topological defects: $t$, $\bar t$ and $\tilde t$. 

The master action we start from is:

\eqn\mdt{\eqalign{ S_{\rm M} = &\int {(-1)^{p-1}\over g^2} \left( 
dB_{p+1} + \bar t \bar t_{p+2} \right) \wedge^*
\left( dB_{p+1} + \bar t \bar t_{p+2} \right) \cr
&+ H_{d-p-1} \wedge 
\left({1 \over e} dA_p + \tilde m B_{p+1} +  t  t_{p+1} \right) 
 \cr &+ {(-1)^{d-p-2}\over e^2} \left( 
H_{d-p-1} + \tilde t \tilde t_{d-p-1} \right) \wedge^* \left( 
H_{d-p-1} + \tilde t \tilde t_{d-p-1} \right)\ .\cr}}
As before we will have two dual forms, $A_p$ and $\tilde A_{d-p-2}$,  with their
respective dual topological defects,
$t_{p+1}$ and $\tilde t_{d-p-1}$.
The $B_{p+1}$ form and its topological defects $\bar t_p$ are not dualized.

Integrating over the form $H_{d-p-1}$ we obtain the compact version of the
massive gauge theory:

\eqn\stc{\eqalign{ S = &\int {(-1)^{p-1}\over g^2} \left( 
dB_{p+1} + \bar t \bar t_{p+2} \right) \wedge^*
\left(dB_{p+1} + \bar t \bar t_{p+2} \right) \cr 
&+ {(-1)^p e^2 \over 4}
\left({1 \over e} dA_p + \tilde m B_{p+1} +  t  t_{p+1} \right) \wedge^* 
\left({1 \over e} dA_p + \tilde m B_{p+1} +  t  t_{p+1} \right) 
 \cr &+ \tilde t \tilde t_{d-p-1} \wedge \left({1 \over e} dA_p + \tilde m B_{p+1}  \right)
 + \tilde t t \tilde t_{d-p-1} \wedge t_{p+1}\ . \cr }}
Here $t_{p+1}$ enters as a singular form,  
due to the compactness of the gauge group, while the $\tilde t_{d-p-1}$ form appears as
a (non-conserved) current minimally coupled to $\left({1 \over e} 
dA_p + \tilde m B_{p+1} \right)$. The last term is a generalized Aharonov-Bohm
interaction between the two topological defects that leads to a generalized Dirac
quantization condition and does not contribute to the partition function \pin.

Integration over the Lagrange multiplier $A_p$ implies that $H_{d-p-1} =
d \tilde A_{d-p-2}$, with an action given by:

\eqn\dbf{\eqalign{\tilde S = &\int {(-1)^{p-1}\over g^2} \left( 
dB_{p+1} + \bar t \bar t_{p+2} \right) \wedge^*
\left(dB_{p+1} + \bar t \bar t_{p+2} \right) \cr 
&+ {(-1)^{d-p-2}  \over  e^2}
\left( d \tilde A_{d-p-2} + \tilde t \tilde t_{d-p-1} \right) \wedge^* 
\left( d \tilde A_{d-p-2} + \tilde  t \tilde t_{d-p-1} \right) 
 \cr &+ \tilde m B_{p+1} \wedge  d \tilde A_{d-p-2} 
 + t  t_{p+1} \wedge d \tilde A_{d-p-2}\ . \cr }}
Here, the dual form $\tilde t_{d-p-1}$ is the one connected to the compactness of the
gauge group, while $ t_{p+1}$ appears as a conserved current ($J_{d-p-2} =
(-1)^{d-1+(p+1)^2} \left(^* dt_{p+1} \right)$) minimally coupled to
$\tilde A_{d-p-2}$.
As we said, the $B_{p+1}$ form does not participate in the duality, and so
the singular form $\bar t_{p+2}$ appears in the same way in both models: as a
singular form due to the compactness of the gauge symmetries \cgs\ and \igs.
We have so established a duality between MGTs and TMGTs in case of a compact gauge
group. The two dual actions are given by the equations \stc\ and \dbf.

By integrating over the tensor field $A_p$ in \stc\ and over
$\tilde A_{d-p-2}$ in \dbf, we obtain an effective action
for the higher rank-tensor $B_{p+1}$. 
In the case of the MGTs described by the action \stc,
it was found in \cri\ that, when the topological defects $t_{p+1}$ are
dilute, this effective action still possesses a massive pole.
In the case in which, instead, these topological defects are in a dense
phase, the effective action we get is:

\eqn\ecb{S_{\rm eff} = \int {(-1)^{p+1}\over g^2} \left( dB_{p+1}
+  \bar t_{p+2} \right) \wedge * \left( dB_{p+1} +  \bar t_{p+2} \right) \ .}
The mass term for $B_{p+1}$ is no longer present:
the condensation of the topological defects $t_{p+1}$ prevents 
the tensor $B_{p+1}$ to become massive through 
the Higgs-St\"uckelberg mechanism.
The same happens in the dual model \dbf\ when we consider the 
effective action for the higher rank-tensor $B_{p+1}$.
Again when the topological defects $t_{p+1}$ are
dilute, this effective action still possesses a massive pole,
while in the case in which these topological defects are in a dense
phase, the effective action is given by \ecb, preventing 
the tensor $B_{p+1}$ to become massive through the topological mechanism.

A crucial role in the determination of the phase diagram is played by the
space-time dimension and by the dimension of the topological defects.
In what follows we will give an example of the effect of the
condensation of the topological defects in the 4-dimensional case.
We will explicitely show, with a separate analysis of the phase structure of the
two models, that they, indeed, admit the same phases.
The generalization to arbitrary dimensions will be briefly discussed at the end.
A more detailed discussion of the possible phases in arbitrary space-time
dimensions, and the study of the conditions for the condensation of
topological defects is left for a forthcoming publication \ref\ffp{M.C.
Diamantini, in preparation.}.

\newsec{Condensation of Topological Defects in the 4-Dimensional Case}

We will now concentrate on the phase structure of the two dual theories
in 4-dimensional Euclidean space-time. We will start with the case $p=1$, and
show at the end that the result for $p=1$ can be generalized to arbitrary rank and
arbitrary space-time dimensions.
For $p=1$ the higher order antisymmetric tensor is  a two form representing
the Kalb-Ramond tensor $B_{\mu \nu}$, while the two lower order tensors are 
the Maxwell field $A_\mu$ and its dual $\tilde A_\mu$.
In order to study the phase structure of the two models we need to introduce an
external probe: since the topologically massive gauge theory \tmm\ can couple
only to closed $p$-branes we will couple the theory to a conserved two form
$J_{\mu \nu}$, that can be interpreted as the worldsheet of a closed string.
The coupling with it plays the role of the (non-local) order parameter, the Wilson
``surface'' $W_S$, for the phase transitions in the theory.
This is the generalization  of the Wilson
loop to objects of one dimension higher. Notice that,
while the standard confinement of point particles is bescribed by an 
area law for the surface enclosed by the worldline of the particles, 
the corresponding phenomenon for  strings is given by a volume law 
for the volume enclosed by the worldsheet of the
strings. 

The two dual actions describing the massive and the topologically massive gauge
theories are:

\eqn\lam{\eqalign{S &= \int d^4x  {1\over 2g^2}
\left( F_{\mu } + \bar t \bar t_{\mu } \right) ^2 + {e^2 \over 4} 
\left( {1\over  e}f_{\mu \nu} + \tilde m B_{\mu \nu} +  t  t_{\mu \nu}
 \right)^2 \cr
&+ i \tilde t  \left( \tilde m B_{\mu \nu} + { 1\over  e} f_{\mu \nu}
\right)\tilde t_{\mu \nu} - ij B_{\mu \nu}  J_{\mu \nu}
\ .\cr}}
and
\eqn\bfz{\eqalign{
\tilde S &= \int d^4x  {1\over 2g^2} \left( F_{\mu } +
 \bar t \bar t_{\mu } \right) ^2 + {1\over e^2} 
\left( h_{\mu \nu} + \tilde t \tilde t_{\mu \nu} \right)^2
+ {i \tilde m}B_{\mu \nu}  \tilde h_{\mu \nu} \cr
 &+ i  t \tilde h_{\mu \nu}  t_{\mu \nu} - 
 ij  B_{\mu \nu}J_{\mu \nu} \ .\cr}}
Here $F_\mu$ is the dual of the Kalb-Ramond field strength, $f_{\mu
\nu}$ is the Maxwell field strength for $A_\mu$ and $h_{\mu \nu}$ 
the Maxwell field strength for $\tilde A_\mu$. The dual of the Maxwell field
strength will be denoted by $\tilde f_{\mu \nu}$ for $A_\mu$ and by
$\tilde h_{\mu \nu}$ for $\tilde A_\mu$.
$g$, $e$ are dimensionless coupling
constants. As we said before, in order to have all gauge symmetries \cgs\ and
\igs\ compact, we need to introduce three different types of topological defects:
$\bar t_{\mu },\ \tilde t_{\mu \nu }$ and  $t_{\mu \nu}$. 
$\bar t_{\mu }$ is associated with the tensor $B_{\mu \nu}$, 
$t_{\mu \nu }$ with $A_\mu$ and  $\tilde t_{\mu \nu}$ with $\tilde A_\mu$.
The last terms in \lam\ and \bfz, represent the coupling to  the Wilson
``surface'' $W_S$.

To properly define the models we use a lattice 
regularization. The lattice we consider is a hypercubic lattice 
with lattice spacing $l$ in four 
Euclidean dimensions , with sites  denoted by $x$
(lattice notation is defined in the Appendix).
The two compact massive and  topologically massive gauge theories to be
 considered will be described
by  partition functions of the Villain type \ref\vil{For a review see: H. Kleinert,
``Gauge Fields in Condesed Matter'', World Scientific, Singapore (1989).}.
With lattice regularization the role of the minimally coupled
topological defects $\tilde t_{\mu \nu }$ in \lam\ and $ t_{\mu \nu}$ in \bfz\
becomes explicit: the sum over these two integer fields in the Villain
 formulation, in the phase in which they condense,  has the effect of breaking the global gauge
symmetries of the two models to a discrete group, $Z_{\tilde t}$ for \lam\ and 
$Z_{t}$  for \bfz\ \ref\card{J.L. Cardy and E. Rabinovici, Nucl.
Phys. B205 (1982) 1.}.

For the theory described in \lam\ the expectation value  
$\langle W_S \rangle$ is given by:

\eqn\top{\eqalign{\langle W_S \rangle &= 
{1\over Z_{\rm top} } \sum_{{\{\tilde t_{\mu \nu }, t_{\mu \nu}\}}
\atop {\{\bar t_{\mu }\}}} {\rm exp } \left( - S_{\rm top} - W_{\rm top} -
W_0 \right) \cr  S_{\rm top} &= \sum_{{\bf
x}, \mu } \ {\pi^2\over 2g^2} \ \left( \bar t_{\mu } - {g\over l  m}
t_{\mu }\right)
 {{m^2 \delta _{\mu \nu} -d_{\mu }\hat d_{\nu }} \over {m^2-\nabla ^2}}
 \left( \bar t_{\nu }
 - {g\over l  m}  t_{\nu }\right) + \cr 
 &+ {\tilde t^2\over 2 e^2}
 \tilde t_{\mu \nu} {O^M_{\mu \nu \alpha \beta}  \over {m^2-\nabla ^2}}
 \tilde t_{\alpha \beta} + {i \pi \tilde t m\over g e}  \bar t_{\mu } 
{K_{\mu \nu \alpha} \over {m^2-\nabla ^2}} \tilde t_{\nu \alpha} +
{i \pi \tilde t\over e l}   t_{\mu } 
{K_{\mu \nu \alpha} \over {m^2-\nabla ^2}} t_{\nu \alpha}\ ,\cr
W_0 &= \sum_{{\bf x}, \mu } {j^2 g^2 \over l^2}  J_{\mu \nu}{1 \over
{m^2-\nabla ^2}} J_{\mu \nu } = \sum_{{\bf x}, \mu } 2 g^2  j^2  V_\mu {M_{\mu \nu} \over
{m^2-\nabla ^2}} V_\nu  \ ,\cr
W_{\rm top} &= \sum_{{\bf x}, \mu } + 2 i \pi j \   \bar t_{\mu } 
{M_{\mu \nu} \over {m^2-\nabla ^2}} V_\nu + {2i \pi j g  \over l} t_\mu
{m \over {m^2-\nabla ^2}} V_\nu
+ {2  t j  m g \over e}
\tilde t_{\mu \nu} {K_{\mu \nu \alpha }  \over {m^2-\nabla ^2}}
 V_{\alpha} \ .\cr}}
Here $l K_{\mu \nu \alpha} V_\alpha = J_{\mu \nu}$, with
$V_\mu$ is the volume enclosed by the surface $J_{\mu \nu}$, $m = \tilde m g e$, 
and we have reabsorbed $\bar t$ and $\tilde t$ into the integer
fields $\bar t_\nu$ and $\tilde t_{\mu \nu}$. 
$t_\mu = l K_{\mu \nu \alpha} t_{\nu \alpha}$ is the physical
integer degree of freedom that describes the topological defects
associated with  the Maxwell gauge field. It
describes closed (or infinitely long) strings of magnetic charge: $\hat
d_\mu  t_\mu = 0$.

For the topologically massive theory \bfz\ the expectation value 
$\langle W_S \rangle$ is, again,
given by:

\eqn\tto{\eqalign{\langle W_S \rangle &= 
{1\over Z_{\rm top} } \sum_{{\{t_{\mu \nu }, \bar t_\mu \}}
\atop {\{\tilde t_{\mu \nu }\}}} {\rm exp } \left( - S_{\rm top} - W_{\rm top} -
W_0 \right) \cr  S_{\rm top} &= \sum_{{\bf
x}, \mu } {2 \pi^2\over e^2} \tilde  t_{\mu \nu }{ O^M_{\mu \nu \lambda \omega} 
\over {m^2-\nabla ^2}}
\tilde t_{\lambda \omega } + {\pi^2\over 2g^2} \  \bar t_{\mu } {{m^2 \delta _{\mu \nu
} -d_{\mu }\hat d_{\nu }} \over {m^2-\nabla ^2}}\bar t_{\nu } + 
{e^2 \tilde t^2 \over 8 l^2}  t_\mu {1 \over {m^2-\nabla ^2}} 
 t_\mu \cr
&+ {2 i\pi^2 m \over eg}\   \bar t_{\mu }{K_{\mu \nu \alpha}\over {m^2-\nabla ^2}}
\tilde t_{\nu \alpha } + {m e  t \over 2 g l}
t_\mu {1 \over {m^2-\nabla ^2}} \bar t_\mu + {i \pi t \over l} 
\tilde t_{\mu \nu} {K_{\mu \nu \alpha}\over {m^2-\nabla ^2}} t_\alpha \ ,\cr
W_0 &= \sum_{{\bf x}, \mu } 2g^2 j^2 V_\mu {M_{\mu \nu} \over {m^2-\nabla
^2}}V_\nu \ , \cr
W_{\rm top} &= \sum_{{\bf x}, \mu } + 2i \pi j \bar t_\mu {M_{\mu \nu} 
\over {m^2-\nabla
^2}}V_\nu  + {4g j m \over e} \tilde t_{\mu \nu} {K_{\mu \nu \alpha}\over 
{m^2-\nabla ^2}} V_\alpha + 
{ij t \pi e g\over 2 l} t_\mu {m  \over {m^2-\nabla ^2}} V_\mu 
\ ,\cr}}
where $J_{\mu \nu} = l K_{\mu \nu \alpha} V_\alpha$, $m = \tilde m g e$, and  $t_\mu = 
l K_{\mu \nu \alpha}  t_{\nu \alpha}$ with
$\hat d  t_\mu = 0$. 
Now, since $ t_\mu$ is minimally coupled to $\tilde  A_\mu$, the role of
 these topological defects as a conserved current of magnetic charges
 is evident: indeed they enter the
theory as an external current coupled to the dual of the electromagnetic field
$A_\mu$. 

The phase structure of \top, in absence of the term $i \tilde t  
\left( \tilde m B_{\mu \nu} + { 1\over  e} f_{\mu \nu}
\right)\tilde t_{\mu \nu}$, was 
discussed in \cri\ (this correspond to the case in which
the topological defects  $\tilde t_{\mu \nu}$
are dilute). It was found that the condensation of $ t_\mu$ leads
to an expectation value of the Wilson surface of the form 

\eqn\nto{\eqalign{\langle W_S \rangle &= {1\over Z_{\rm top} } 
\sum_{\{ \bar t_\mu\} }
 {\rm exp } \left( - S_{\rm top} - W_{\rm top} -
W_0 \right) \cr  S_{\rm top} &= \sum_{{\bf
x}, \mu } \ - {\pi^2\over 2g^2} \  Q {1 \over
\nabla ^2} Q - {t^2 \over e^2}\tilde  t_{\mu \nu} \tilde t_{\mu \nu} \ ,\cr
W_0 &= \sum_{{\bf x}, \mu } - {g^2  j^2\over l^2}\  J_{\mu \nu}
{1 \over \nabla ^2} J_{\mu \nu }\ ,\cr
W_{\rm top} &= \sum_{{\bf x}, \mu } - {i \pi j\over l}\  \bar t_\mu
{\hat K_{\mu \nu \alpha} \over \nabla ^2} J_{ \nu \alpha }
 \ ,\cr}}
where $Q = l \hat d_\mu \bar t_\mu$. $\bar t_\mu$ 
and $t_\mu$ are both connected with the compactness of the gauge symmetry \cgs, 
and, thus, they enter the theory on the same footing, as magnetic strings.
$Q$ are the monopoles that live at
the end-points of the strings $\bar t_\mu$ .
In this magnetic condensation phase, a lenghty calculation shows that 
the self energy of a circular charge loop 
of radius $R$ is proportional to $R\ln R$, and this gives rise to logarithmic
confinement of electric strings (logarithmic confinement phase)\ref\jja{M.C. 
Diamantini, P. Sodano and C.A. Trugenberger, 
Nucl. Phys. B474 (1996) 641.}.
The monopoles $Q$ are always in a plasma phase :
their condensation (after the condensation of $ t_\mu$), as it was shown
in \cri, gives a volume law and thus the usual 
confinement strings.

Let us see what is the effective action induced by the condensation of 
$t_{\mu }$ in the dual model \bfz.
Also in this case the expectation value of the Wilson ``surface'' is: 

\eqn\ntt{\eqalign{\langle W_S \rangle &= {1\over Z_{\rm top} } 
\sum_{\{ \bar t_\mu\} }
 {\rm exp } \left( - S_{\rm top} - W_{\rm top} -
W_0 \right) \cr  S_{\rm top} &= \sum_{{\bf
x}, \mu } \ - {\pi^2\over 2g^2} \  Q {1 \over
\nabla ^2} Q - {4 \pi^2\over e^2} \tilde  t_{\mu \nu} \tilde t_{\mu \nu} \ ,\cr
W_0 &= \sum_{{\bf x}, \mu } - {g^2  j^2\over l^2}\  J_{\mu \nu}
{1 \over \nabla ^2} J_{\mu \nu }\ ,\cr
W_{\rm top} &= \sum_{{\bf x}, \mu } - {i \pi j\over l}\  \bar t_\mu
{\hat K_{\mu \nu \alpha} \over \nabla ^2} J_{ \nu \alpha }
 \ ,\cr}}
with $Q = l \hat d_\mu \bar t_\mu$ (the difference in the coefficients of
$\tilde t_{\mu \nu}$ is due to the way the topological defects enter the Villain
formulation in the two dual models).
Also in this  phase the self energy of a circular  loop 
of radius $R$ is proportional to $R\ln R$, and this gives rise to logarithmic
confinement of  electric strings: in this dual theory the
objects that condense enter as a minimally coupled current and, therefore,
one can look at this electric
confinement phase as  a ``magnetic Higgs phase''.
It is interesting to notice that in \bfz\ what lead to confinement is the
condensation of the topological defects that enter the theory as a minimally
coupled current and, as we explained before, are associated on the lattice with
the  breaking the global gauge symmetry
down to $Z_{ t}$. A similar phase was found in \jja, in the context of effective field theories for
3-dimensional Josephson junction arrays, studying a model in which the original
gauge group was non-compact and only the BF coupling was periodic.
Also in \ntt, as expected, the further condensation of the $\bar t_\mu$ gives a volume law for the
expectation value of the Wilson ``surface''.

Let us now analyze the effect of the condensation of $\tilde t_{\mu \nu}$. In both
models this leads to an effective action for the Wilson surface of the type
\eqn\its{W_0 = \sum_{{\bf x}, \mu }- \alpha
J_{\mu \nu} J_{\mu \nu} \ ,}
with $\alpha = j^2 g^2$ .
This term measures the area of the Wilson ``surface''. This Wilson surface
can be seen as the world-sheet of a closed string and \its\ is just the standard
Nambu-Goto action for a string with string tension ${1\over \alpha}$.
\its\ correspond to an area law for surfaces and it does thus not describe any
type of confinement for strings.

From this we learn that both in MGT and TMGT only the condensation of one type
of topological defects, $t_\mu$, leads to a confinement phase, while the
condensation of the dual topological defects $\tilde t_{\mu \nu}$ has only the 
effect of promoting the effective action for the probe surface to a Nambu-Goto type of
action. The dual phase, in which magnetic charges are confined, is absent.

In \cri\ it was proved that the confinement phase we found in the 4-dimensional
case and for rank $p=1$ for the MGTs described by the action \stc,
 is present also in arbitrary space-time dimensions and
for arbitrary $p$ ($p \le d-1$). Simply invoquing the duality we know that the
phase is present also in the TMGTs, described by the action \dbf, in arbitrary
space-time dimensions and for arbitrary rank antysymmetric tensor fields.
The role of the condensation of $\tilde t_p$ will be discussed in \ffp.

\newsec{Conlusions}

In this paper we have  shown that the duality between MGTs and TMGTs \spa\ can
be extended also to include topological defects. We have also explicitly
shown that the condensation of these
defects leads, in 4-dimensions, to  confinement  for electric 1-branes in the two dual models.
This phase can be seen as a confinement phase in one model and as a 
Higgs phase in the dual model in which the topological defects that condense
enter as a minimally coupled current.
The dual phase in which magnetic branes are confined is absent.
The presence of the confinement phase explicitely found in the 4-dimensional
case, can be easily generalized, using the
duality between \stc\ and \dbf, to arbitrary space-time dimensions.

\newsec{Appendix}
On the lattice, we define the following forward and backward 
derivatives and shift operators: 

\eqn\der{\eqalign{d_\mu f({\bf x}) &\equiv {f({\bf x} + \hat \mu l) -
f({\bf x}) \over l}\ , \qquad S_\mu f({\bf x)} \equiv f({\bf x} +
\hat \mu l) \ , \cr
\hat d_\mu f({\bf x}) &\equiv {f({\bf x}) - f({\bf x} - \hat \mu l)
\over l} \ , \qquad \hat S_\mu f({\bf x}) \equiv f({\bf x} - \hat \mu
l) \ . \cr}}
Summation by parts interchanges the two derivatives, with a minus sign,
and the two shift operators.
We also introduce the three-index lattice operators \jja:

\eqn\ddk{K_{\mu \nu \alpha} = S_\mu S_\nu \epsilon_{\mu \phi \nu \alpha}
d_\phi \ , \quad \hat K_{\mu \nu \alpha} =  \epsilon_{\mu \nu
 \phi \alpha} \hat d_\phi \hat S_\alpha \hat S_\nu \  .}
These operators are gauge-invariant in the sense
that:

\eqn\pdk{\eqalign{K_{\mu \nu \alpha} d_\alpha &= K_{\mu \nu \alpha}
d_\nu = \hat d_\mu K_{\mu \nu \alpha} = 0 \ , \cr
\hat K_{\mu \nu \alpha} d_\alpha &= \hat d_\mu \hat K_{\mu \nu
\alpha} = \hat d_\nu \hat K_{\mu \nu \alpha} = 0\ .\cr}}
Moreover they satisfy the equations:

\eqn\ker{\eqalign{&\hat K_{\mu \nu \alpha} K_{\alpha \lambda \omega} =
K_{\mu \nu \alpha} \hat K_{\alpha \lambda \omega} = O_{\mu \nu
\lambda \omega } =\cr
&= - \left( \delta_{\mu \lambda} \delta_{\nu \omega} - 
\delta_{\mu \omega} \delta_{\nu \lambda}\right) \nabla^2 + \left( 
\delta_{\mu \lambda} d_\nu \hat d_\omega - \delta_{\nu \lambda}
d_\mu \hat d_\omega \right) - \left( 
\delta_{\nu \omega} d_\mu \hat d_\lambda - \delta_{\mu \omega}
d_\nu \hat d_\lambda \right) \ ,\cr
&\hat K_{\mu \omega \alpha} K_{\omega \alpha \nu } =
K_{\mu \omega \alpha} \hat K_{\omega \alpha \nu } =  2 M_{\mu \nu} = - 
2 \left(  \delta_{\mu \nu} \nabla^2 - d_\mu \hat d_\nu \right) \ . \cr}}
The expressions $O_{\mu \nu \lambda \omega }$ and $M_{\mu \nu}$ are 
lattice versions of the Kalb-Ramond and Maxwell kernels, respectively,
and $\nabla^2 = d_\mu \hat d_\mu = \hat d_\mu d_\mu$ is the lattice
Laplacian. 
We also define the operator 
\eqn\mke{\eqalign{O^M_{\mu \nu
\lambda \omega } &=
 + \left( \delta_{\mu \lambda} \delta_{\nu \omega} - 
\delta_{\mu \omega} \delta_{\nu \lambda}\right) m^2 - \left( 
\delta_{\mu \lambda} d_\nu \hat d_\omega - \delta_{\nu \lambda}
d_\mu \hat d_\omega \right) \cr &+ \left( 
\delta_{\nu \omega} d_\mu \hat d_\lambda - \delta_{\mu \omega}
d_\nu \hat d_\lambda \right) \ .\cr}}

\bigbreak\bigskip\bigskip

\listrefs
\end